\PassOptionsToPackage{numbers,sort&compress}{natbib} 
\documentclass[preprint,  titlepage, 
onecolumn, amsmath,amssymb,aps,superscriptaddress,
nofootinbib]{revtex4-2}

\usepackage{graphicx}
\usepackage{dcolumn}
\usepackage{bm}

\usepackage{float}

\usepackage{hyperref}

\begin{document}

\title{NANOGrav 15-year gravitational-wave signals from binary supermassive black-holes seeded by primordial black holes, {\bf and implications for the origins of Little Red Dots} }

\author{Mikage U. Kobayashi}
\affiliation{Particle and Nuclear Physics Program, Graduate University for Advanced Studies (SOKENDAI), 1-1 Oho, Tsukuba, Ibaraki 305-0801, Japan}
\affiliation{Theory Center, IPNS, High Energy Accelerator Research Organization (KEK), 1-1 Oho, Tsukuba, Ibaraki 305-0801, Japan}

\author{Kazunori Kohri}
\affiliation{Division of Science, National Astronomical Observatory of Japan, and SOKENDAI, 2-21-1 Osawa, Mitaka, Tokyo 181-8588, Japan}
\affiliation{Department of Astronomy, University of Tokyo, Bunkyo-ku, Hongo, Tokyo 113-0033, Japan}
\affiliation{Theory Center, IPNS, High Energy Accelerator Research Organization (KEK),
1-1 Oho, Tsukuba, Ibaraki 305-0801, Japan}
\affiliation{Kavli IPMU (WPI), UTIAS, The University of Tokyo, Kashiwa, Chiba 277-8583, Japan}

\date{\today}

\begin{abstract}
  In this paper, we explain the recently reported a nHz-band
  gravitational-wave background from NANOGrav 15-year through the merger of binary
  super-massive black holes with masses of $10^9 M_{\odot}$ formed by
  the growth of primordial black holes. When a primordial black hole
  accretes at a high accretion rate, it emits a large number of
  high-energy photons. These heat the plasma, causing high-redshift
  cosmological 21cm line emission. Since this has not been detected,
  there is a strict upper bound on the accretion rate. We have found
  that with the primordial black hole abundance
  $10^{-14} \lesssim f_{\rm PBH} \lesssim 10^{-12}$ and the mass
  $1 M_{\odot} \lesssim m_{\rm PBH} \lesssim 10^3 M_{\odot}$, we
  successfully fit the nHz band gravitational wave background from
  NANOGrav 15-year while avoiding the 21 cm line emission. {\bf In addition, we also
  discuss the implication for the origins of the Little Red Dots.} We propose
  that future observations of the gravitational wave background and
  the cosmological 21cm line can test this scenario.
\end{abstract}


\maketitle

\section{Introduction}

Recently the Pulsar Timing Array (PTA) collaborations, including
NANOGrav (North American Nanohertz Observatory for Gravitational
Waves), reported evidence for the presence of a stochastic
gravitational-wave background (GWB)~\cite{NANOGrav:2023gor, Goncharov:2021oub, Reardon:2023gzh, Zic:2023gta, EPTA:2023fyk, EPTA:2023sfo, Xu:2023wog, Antoniadis:2022pcn}. This
observed signal is more than seven orders of magnitude larger than the
GWB predicted by the normal models of the primordial
inflation accreted in the early Universe~\cite{Lyth:2009imm}. Several
scenarios have been proposed for the origin of such a signal. One of
the most promising candidates for the nanohertz (nHz)
gravitational-wave background in astrophysics is a merger of binary
supermassive black holes (SMBHs).~\footnote{
  The other possible sources in the early Universe based on new
  physics beyond the standard model include quantum fluctuations
  during inflation~\cite{Guzzetti:2016mkm}, the scalar-induced
  gravitational
  waves~\cite{Kohri:2020qqd,Kohri:2018awv,Inomata:2019zqy,Inomata:2019ivs,Domenech:2021ztg,Balaji:2023ehk,NANOGrav:2023hvm,Franciolini:2023pbf,Wang:2023ost,Inomata:2023zup,Harigaya:2023pmw,Inomata:2023drn,Inui:2024fgk,Liu:2023ymk, Yang:2025vsq},
  and the collapse of topological defects such as cosmic strings and
  domain
  walls~\cite{Ellis:2023tsl,Kitajima:2023vre,Wang:2023len,Lazarides:2023ksx,Eichhorn:2023gat,Chowdhury:2023opo,Servant:2023mwt,Antusch:2023zjk,
    Yamada:2023thl,Ge:2023rce,Basilakos:2023xof,Kitajima:2023cek,Guo:2023hyp,Blasi:2023sej,Gouttenoire:2023ftk,Barman:2023fad,Lu:2023mcz,Li:2023tdx,Du:2023qvj,Babichev:2023pbf,Gelmini:2023kvo},
  as well as cosmological phase
  transitions~\cite{Fujikura:2023lkn,Addazi:2023jvg,Bai:2023cqj,Megias:2023kiy,Han:2023olf,Zu:2023olm,Ghosh:2023aum,Xiao:2023dbb,Li:2023bxy,DiBari:2023upq,Cruz:2023lnq,Gouttenoire:2023bqy,Ahmadvand:2023lpp,An:2023jxf,Wang:2023bbc, Ashoorioon:2022raz}.
}  
The SMBHs with masses $10^6 M_\odot\lesssim m$,
are believed to reside at the centers of most of galaxies
\cite{Kormendy:2013dxa, EventHorizonTelescope:2019dse}.  Within the
hierarchical clustering scenario of structure formation in the
$\Lambda$CDM cosmology, dark halos of cold dark matter (CDM) grow
through gravitational clustering and successive mergers. Galaxies form
within halos through gas cooling, star formation, and feedback
processes, and galaxy mergers naturally follow from the halo
mergers. In this context, the SMBHs residing in the centers of merging
galaxies are expected to sink toward the center of the remnant galaxy
and form binary SMBHs~\cite{Begelman:1980vb}. Interactions with stars
and gas extract energy and angular momentum from the system, reducing
the orbital separation of the binary. Once the separation reaches
sub-parsec scales, gravitational-wave emission dominates the energy
loss, driving the inspiral and ultimately leading to coalescence of
the SMBHs. During the inspiral phase, the binary SMBHs radiate
gravitational waves in the nHz frequency band of the
PTA~\cite{Rajagopal:1994zj,Jaffe:2002rt,Wyithe:2002ep}.

However, even in the latest numerical
simulations~\cite{Enoki:2004ew,Sesana:2008mz}, the theoretical GW
signal from binary SMBH mergers is pointed out to be a factor of a few
or even more smaller than the reported NANOGrav 15-year GW signal. It
is highly ironic that this model, considered the most natural
astrophysical model, cannot explain the NANOGrav 15-year GW signal.

Therefore, in this paper, we propose primordial black holes (PBHs)
formed in the early Universe as seed BHs~\cite{Carr:2020gox}. The PBHs
could be produced by the gravitational collapse of high density
fluctuation at a small scale, which is predicted, e.g., in some class
of inflation models~\cite{Carr:2020gox}.  The seed PBHs grow via
accretion into supermassive black holes (SMBHs) with masses
$m \sim 10^9 M_{\odot}$~\cite{Kohri:2022wzp}.  Of course, this model
is classified as one based on new physics beyond the Standard
Model. These SMBHs originating from the seed PBHs reside at the
centers of high-redshift galaxies, just like the SMBHs of the
astrophysical origin.  Furthermore, we propose a scenario where SMBH
mergers occur during galaxy mergers, emitting nHz-band gravitational
waves, fitting the signal detected by NANOGrav 15-year.

In fact, there is another serious problem in astrophysics: the origin
of the high-redshift SMBHs with masses $m \sim 10^9 M_\odot$ observed
up to high redshifts $z \sim 7$ remains unknown.
Although several astrophysical mechanisms have been proposed, forming SMBHs at high redshift remains a challenging and not yet fully understood problem~\cite{Willott:2010yu, 2021ApJ...907L...1W}. In
fact, to resolve this problem, the scenario where the seed PBHs
accreted to form high-redshift SMBHs is highly desirable.  However, it
is not simply a matter of accreting at a high accretion rate.

If an accretion disk forms at such an Eddington or super-Eddington
accretion rate, it will emit an enormous amount of high-energy
photons. This mechanism must be consistent with checks from other
cosmological observations.  Such high-energy photon emissions before
the reionization epoch of the Universe (at a redshift $z > 10$) should
have heated the gas and/or plasma in the Universe, affecting the
absorption/emission of the cosmological 21cm
line~\cite{Kohri:2022wzp}.  For example, the observation that no
cosmological 21cm emission line was found at $z \sim 17$ provides a
stringent upper bound on the accretion on to the seed
PBHs~\cite{Bowman:2018yin,Singh:2021mxo}.  Only with satisfying this
constraint, the seed PBHs can evolve into the SMBHs until
$z \sim 7$~\cite{Kohri:2022wzp}. Then, the high-redshift binary SMBHs
thus emit nHz-band GWs during their mergers, fitting the NANOGrav
15-year GW signal.  In this paper, we discuss this scenario and examines
the required mass and abundance of the seed PBHs, which can be also
the origin of the high-redshift SMBHs.

In addition, objects known as Little Red Dots, recently discovered by JWST, have been suggested to represent a transitional phase of mature active galactic nuclei (AGNs)~\cite{inayoshi2025critical}.
In Ref.~\cite{DeLuca:2025nao}, it has been suggested that the transitional phase in which black holes grow into supermassive black holes (SMBHs) via accretion onto primordial black hole (PBH) seeds may be observed as Little Red Dots. In this context, in our scenario as well, the transitional phase before reaching $10^9 M_\odot$ SMBHs can be interpreted as being observed as Little Red Dots.

The structure of the paper is as follows: In Section II, we outlines
the current status of the GW signal emitted from the merging binary
SMBHs and compare it with the observation by the NANOGrav 15-year. In
Section III, we explain the method about how to theoretically
calculate the GW signal emitted by merging binary SMBHs in detail. In
Section IV, we show the main results of this paper with considering
the binary SMBHs originated from PBHs. We summarize our findings and
conclude in Section V. Throughout this paper, we adopt natural units
where $\hbar = c = 1$.

\section{NANOGrav 15-year and Binary Supermassive Black-Holes}

\begin{figure}[!h]
  \centering
  \includegraphics[width=0.8\columnwidth]{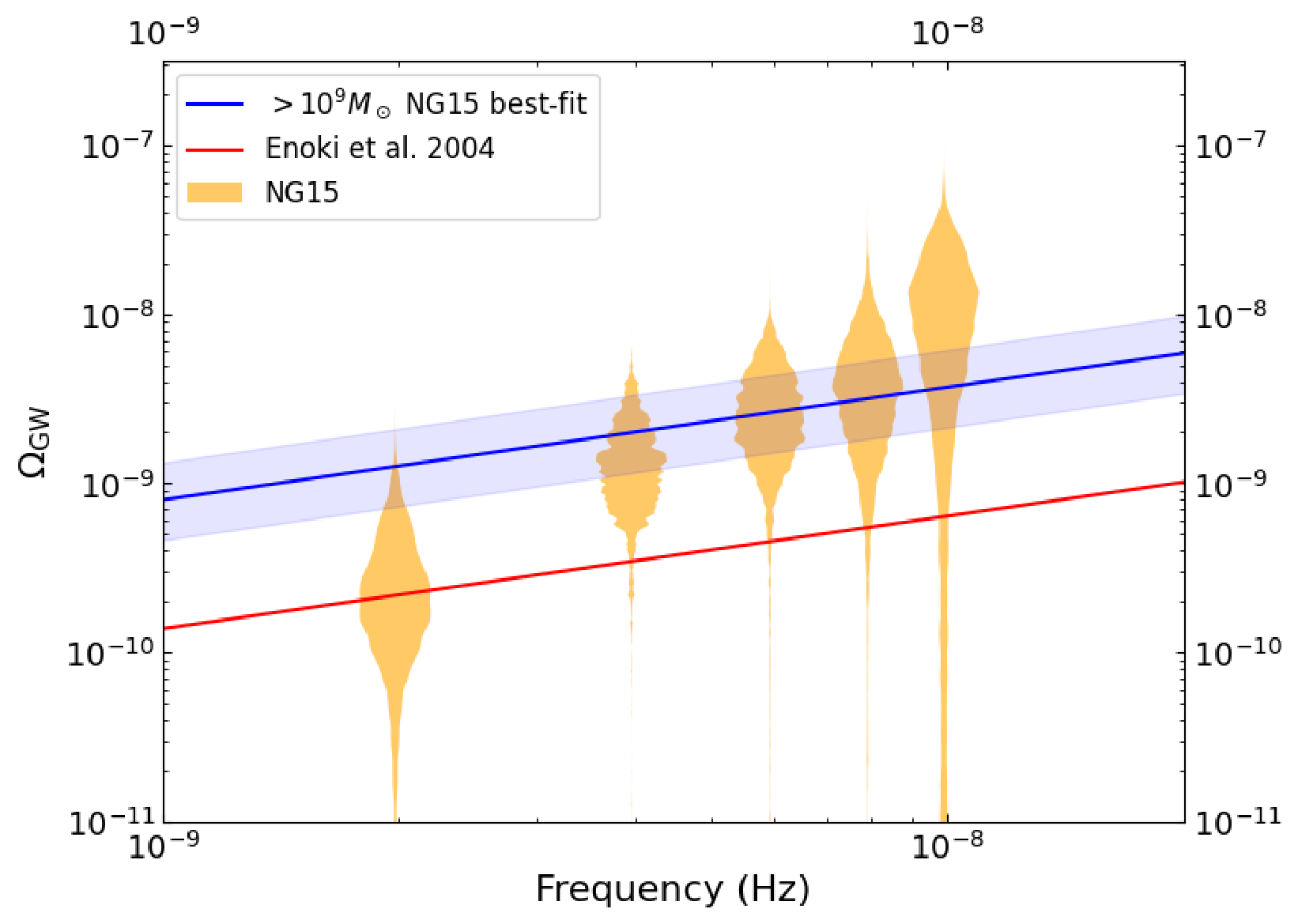}
  \caption{Comparison of the best-fit GW spectrum of the        NANOGrav
      15-year (NG15) under the scenario of the binary SMBH merger
      (blue solid line: median; shaded region: 90\% credible interval)
      with the prediction by the numerical simulation (red solid
      line)~\cite{Enoki:2004ew}. The horizontal axis indicates the
      frequency, and the vertical axis represents the mean GW energy
      density spectrum defined in Eq.~\ref{eq:Omega_GW2}. The first
      five data points of the NG15 are presented as the violin plots.}
    \label{fig:NG15}
\end{figure}

Assuming the characteristic spectral shape $f^{-2/3}$ as a function of
the frequency $f$ appropriate for the binary SMBH inspirals, the
analysis of the NANOGrav 15-year shows that the strain amplitude is
measured to be $A=2.4^{+0.7}_{-0.6} \times10^{-15}$ (median and 90\%
credible interval) at
$f =  1 \mathrm{yr}^{-1} \simeq 3 \times 10^{-8}\mathrm{Hz}$~\cite{NANOGrav:2023gor}.  On the
other hand, the numerical simulations predict a strain amplitude of
$A \simeq 1 \times 10^{-15}$ at
$f \simeq 3 \times
10^{-8}\mathrm{Hz}$~\cite{Enoki:2004ew}~\footnote{
  See also the results of the numerical simulation performed by Sesana
  et al.~\cite{Sesana:2008mz}.}.
This estimate corresponds to the most optimistic case, in which every
galaxy merger inevitably leads to the coalescence of the central
SMBHs, and yet it still falls short of the signal of the NANOGrav
15-year by a factor of several. In Fig.~\ref{fig:NG15}, we compares
the prediction of the numerical simulation with the signal of the
NANOGrav 15-year. The blue region represents the spectrum inferred
from fitting the strain amplitude of the NANOGrav 15-year (violin
plot), while the red line shows the prediction of the numerical
simulation. The comparison reveals that the simulated amplitude is
smaller by a factor of several than the best-fit value of the
NANOGrav.

In this work, we investigate whether this discrepancy can be
alleviated by an additional contribution from the SMBHs of a
primordial origin. Specifically, we consider the scenario in which
primordial black hole (PBH) formed in the early Universe grows through
accretion on a seed black hole into the SMBH that contribute to the
nanohertz gravitational-wave background. Since the SMBHs with masses
$m \gtrsim10^9 M_\odot$ are the main contributors to the
gravitational-wave signal at the nHz band, we focus on the case where
light PBH seeds grow to $m \sim10^9 M_\odot$. We consider the range of
the initial PBH seed mass to be
$M_{\rm PBH} \lesssim 10^4 M_\odot$~\cite{Carr:2020gox} to escape from
the limits by the $\mu-$distortion of the cosmic microwave
background (CMB)~\cite{Kohri:2014lza}, and the accoustic
reheating~\cite{Jeong:2014gna,Inomata:2016uip}. The accretion-driven
growth of the PBHs is subject to constraints from observations of the
high-redshifted cosmological 21cm line~\cite{Kohri:2022wzp}. So far
any emission lines of the high-redshifted cosmological 21cm line has
not been observed~\cite{Bowman:2018yin,Singh:2021mxo}, which gives the
upper bound on the number density of the BH accretion system with the
Eddington or super-Eddington accretion rates evolving to a
SMBH. According to Ref.~\cite{Kohri:2022wzp}, if the comoving number
density of the seed PBHs could be
$n_{\rm PBH}=10^{-5}\mathrm{Mpc}^{-3}$, the initial mass of the seed
PBH must satisfy the upper bound on the masses,
$M_{\rm PBH}\lesssim10^4 M_\odot$ in order to grow to
$m \sim10^9 M_\odot$ until the redshift $z \sim 7$. Similarly, for
$n_{\rm PBH}=10^{-3}\mathrm{Mpc}^{-3}$, the constraint is given to be
$M_{\rm PBH}\lesssim 10^2 M_\odot$.  Motivated by these
considerations, we examine the parameter range of the comoving number
density of the seed PBHs
$n_{\rm PBH}=10^{-5}\mathrm{Mpc}^{-3}$–$10^{-3}\mathrm{Mpc}^{-3}$,
with the masses $1M_\odot$ – $10^3M_\odot$ formed by $z \gtrsim 30$,
that subsequently grow to $m \sim10^9 M_\odot$ until the redshift
$z \sim 7$. These number density might be higher than the ones
observed at around $z \sim 7$~\cite{Willott:2010yu,Kohri:2022wzp}.
However, here we interpret that most SMBHs are not emitting photons as
AGNs and remain undiscovered in high-redshift optical and/or infrared
observations, where observational precision has only recently
improved.  We further consider that the accretion ceases after around
$z \sim 7$ for some reasons, leaving the PBH-origin SMBHs in the
centers of galaxies that then evolve passively as well as the normal
high-redshifted SMBHs in the standard scenarios.  We investigate how
the inclusion of such a PBH-origin SMBH population in addition to the
standard SMBH abundance modifies the GW energy density spectrum.

\section{Method}

\subsection{Gravitational Wave Spectrum from Binary SMBH Mergers}
The mean gravitational wave (GW) energy-density spectrum emitted from
the binary SMBH population is evaluated to be~\cite{Phinney:2001di}
\begin{equation}
    \label{eq:Omega_GW2}
    \begin{aligned}
        \Omega_{\rm GW}(f) = \frac{1}{\rho_c} \frac{\mathrm{d}\rho_{\rm GW}}{\mathrm{d} \ln{f}} = \frac{2\pi}{5G} \int \mathrm{d}m_1 \mathrm{d}m_2 \frac{\mathrm{d}z}{1+z} \frac{\mathrm{d}R_{\mathrm{BH}}}{\mathrm{d}m_1 \mathrm{d}m_2} \frac{\mathrm{d}V_c}{\mathrm{d}z} \frac{f^3 |\tilde{h}(f)|^2}{\rho_c} 
    \end{aligned}
\end{equation}
where $G$ is the gravitational constant, $\mathrm{d}V_c/\mathrm{d}z$
is the differential comoving volume, $\rho_{GW}$ is the present GW
energy density, $\rho_c$ is the critical density of the Universe,
$m_i$ denotes each BH mass ($i$ = 1 or 2) in the binary, and $f$ is
the GW frequency.  
In this study, asymmetric binaries with mass ratios different from unity are taken into account in the integration over the mass distribution. On the other hand, highly asymmetric systems such as $10^7 + 10^9 M_\odot$, where one component is lighter than $10^9 M_\odot$, as well as binaries composed of light PBHs, are not included in our calculation. This is because the gravitational-wave amplitude strongly depends on the chirp mass, which becomes smaller for such systems, leading to a significant suppression of their contribution. In particular, mergers of light PBHs mainly contribute to higher frequency bands, and their impact on the nHz band is negligible. As a result, the gravitational-wave background in the nHz band considered in this work is confirmed to be dominated by mergers of SMBHs with masses above $10^9 M_\odot$, while other contributions are subdominant.
The amplitude $|\tilde{h}(f)|$ is modeled using the
inspiral-merger-ringdown template proposed in~\cite{Ajith:2007kx}:
\begin{equation}\notag
    |\tilde{h}(f)| = \sqrt{\frac{5}{24}}\frac{{(G\mathcal{M}_z)}^{\frac{5}{6}}}{\pi^{\frac{2}{3}}D_L} \\ \times
    \begin{cases}
        f^{-\frac{7}{6}} & f < f_{\mathrm{merg}} \\
        f_{\mathrm{merg}}^{-\frac{1}{2}}f^{-\frac{2}{3}} & f_{\mathrm{merg}} \leq f < f_{\mathrm{ring}} \\
        f_{\mathrm{merg}}^{-\frac{1}{2}}f_{\mathrm{ring}}^{-\frac{2}{3}}\frac{\sigma^2}{(4(f-f_{\mathrm{ring}})^2 + \sigma^2)} & f_{\mathrm{ring}} \leq f < f_{\mathrm{cut}}.
    \end{cases}
\end{equation}
Here, $\mathcal{M}_z = (1 + z)(m_1 + m_2)\eta^{\frac{3}{5}}$ is the
redshifted chirp mass, $\eta=m_1m_2/(m_1 + m_2)^2$ is the symmetric
mass ratio, $D_L$ is the luminosity distance, $f_{\mathrm{merg}}$ is
the merger frequency, $f_{\mathrm{ring}}$ is the ringdown frequency,
$f_{\mathrm{cut}}$ is the cutoff frequency, and $\sigma$ is the width
of the ringdown phase. The frequencies $f_{\mathrm{merg}}$,
$f_{\mathrm{ring}}$, $f_{\mathrm{cut}}$, and $\sigma$ are
parameterized as
\begin{equation}
    f_i = \frac{\eta^{\frac{5}{3}}}{\pi G \mathcal{M}_z} (a_i\eta^2+b_i\eta+c_i),
\end{equation}
$a_i$, $b_i$, and $c_i$ are fitting coefficients, with their values listed in Table I of Ref.\cite{Ajith:2007kx}.
Since we
focus on the GWB at the nanohertz bands in this study, we
note that the dominant contribution arises from the inspiral phase.
The BH merger rate ${\rm d}R_{\rm BH}/{\rm d}m_1 {\rm d}m_2$ is
evaluated following~\cite{Ellis:2023owy}. Within the Extended
Press-Schechter (EPS) formalism, the merger rate of halos is computed,
assuming the following halo mass–BH mass
relation~\cite{Barkana:2000fd,Wyithe:2002ep}:
\begin{equation}\label{eq:halo-BH}
    M = 10.5 \times 10^{12}  M_{\odot} \left[\frac{\Omega_M(0)}{\Omega_M(z)}\frac{\Delta_c(z)}{18\pi^2}\right]^{-\frac{1}{2}} (1 + z)^{-\frac{3}{2}} \left(\frac{m}{10^8 M_{\odot}}\right)^{\frac{3}{5}},
\end{equation}
with the halo mass $M$. Here, $\Omega_M(z)$ is the matter density
parameter at a redshift $z$, and
$\Delta_c(z)=18\pi^2+82[\Omega_M(z)-1]-39[\Omega_M(z)-1]^2$ is the
virial overdensity.  Using this relation, the BH merger rate
$R_{\rm BH}$ can be written as
\begin{equation}\label{eq:BH_merger_rate2}
    \begin{aligned}
        \frac{\mathrm{d}R_{\mathrm{BH}}}{\mathrm{d}m_1 \, \mathrm{d}m_2}
        \;\approx\; C_{\mathrm{BH}} \,
        \frac{\mathrm{d}p(M_1, M_2, t)}{\mathrm{d}t \, \mathrm{d}M_2} \,
        \frac{\mathrm{d}M_2}{\mathrm{d}m_2} \,
        \frac{\mathrm{d}n(m_1, t)}{\mathrm{d}m_1}. \
    \end{aligned}
\end{equation}
We compute the merger rate based on the Extended Press–Schechter (EPS) formalism. The EPS framework depends on the halo mass function and its time evolution, and is known to tend to underestimate halo masses at high redshift. However, since mergers at $z \sim \mathcal{O}(1)$ are dominant in this study, the impact of these uncertainties on our final results is expected to be small.
We also assume that SMBHs formed from PBHs follow the BH–halo mass relation~\cite{Barkana:2000fd,Wyithe:2002ep}; however, the validity of this assumption is not necessarily well established. In the case of PBH seed origins, the evolution may differ from that in standard galaxy formation scenarios, which could introduce uncertainties in the normalization and mass dependence of the scaling relation. These differences can directly affect the estimated BH merger rate and, consequently, the amplitude of the GWB.
The coefficient $C_{\rm BH}$, originally introduced
in~\cite{Ellis:2023owy}, represents the probability that BHs merge
when their host halos merge. In this work, however, it is treated as a
fitting parameter, since our aim is to incorporate the contribution of
the SMBHs grown from the PBH seeds into the results of the numerical simulation.
The quantity ${\rm d} p/{\rm d}t{\rm d}M_2$ denotes the probability,
within the EPS formalism, that a halo of mass $M_1$ merges with
another halo of mass $M_2$ to form a halo of total mass
$M_f = M_1 + M_2$ per unit time at a given redshift. It is given by
\begin{equation}
    \begin{aligned}
        \frac{\mathrm{d}p(M_1, M_2, t)}{
        \mathrm{d}t \, \mathrm{d}M_2}
        = \frac{1}{M_f} \sqrt{\frac{2}{\pi}} 
        \left| \frac{\dot{\delta}_c}{\delta_c} \right|
        \frac{\mathrm{d} \ln \sigma}{\mathrm{d} \ln M_f}
        \left[ 1 - \frac{\sigma^2(M_f)}{\sigma^2(M_1)} \right]^{-\tfrac{3}{2}}\frac{\delta_c}{\sigma(M_f)}
   \exp\!\left[ -\frac{\delta_c^2}{2}
   \left( \frac{1}{\sigma^2(M_f)} - \frac{1}{\sigma^2(M_1)} \right) \right].
    \end{aligned}
\end{equation}
Here, $\sigma(M)^2$ is the variance of density fluctuations on the
mass scale $M$, and $\delta_c(z)$ is the critical density contrast for
collapse, approximated by $\delta_c(z) \simeq 1.686/D(z)$, with $D(z)$
being the linear growth function~\cite{Dodelson:2003ft}.  Finally,
${\rm d} n/{\rm d} m(m, t)$ represents the BH mass function. In this
study, the BH mass function is constructed from the EPS-based BH mass
function, supplemented by an additional contribution from the SMBHs
formed through the growth of the PBH seeds.

\subsection{BH Mass Function Including SMBHs from PBH Seeds}

The contribution of the SMBHs originating from the PBH seeds to the GW
energy density spectrum is incorporated as follows.  We add a PBH seed
term to the BH mass function derived from the EPS formalism:
\begin{equation}\label{eq:fullBHMF}
    \begin{aligned}
        \frac{{\rm d}n}{{\rm d}m} = \left.\frac{{\rm d}n}{{\rm d}m}\right|_{\mathrm{EPS}} + \left.\frac{{\rm d}n}{{\rm d}m}\right|_{\mathrm{PBH}}.
    \end{aligned}
\end{equation}

Within the EPS formalism, the halo mass function is given by
\begin{equation}
    \begin{aligned}
       \frac{{\rm d}n}{{\rm d}M}(M,z) 
        = \sqrt{\frac{2}{\pi}} \, \frac{\rho_0}{M^2} 
            \frac{\delta_c(z)}{\sigma(M)} 
            \left| \frac{{\rm d} \ln \sigma(M)}{{\rm d} \ln M} \right|
            \exp\left( - \frac{\delta_c^2(z)}{2 \sigma^2(M)} \right).
    \end{aligned}
\end{equation}
As expressed in Eq.~(\ref{eq:halo-BH}), we assume a one-to-one
correspondence between the halo mass and the BH mass. Using this
relation, the BH mass function derived from the EPS formalism becomes
\begin{equation}
    \begin{aligned}
        \left.\frac{{\rm d}n}{{\rm d}m}\right|_{\mathrm{EPS}}
        = \frac{{\rm d}n}{{\rm d}M} \frac{{\rm d}M}{{\rm d}m}.
    \end{aligned}
\end{equation}

The BH mass function originating from the PBH seeds is modeled using a
log-normal distribution, assuming that the PBH seeds can grow into the
SMBHs with the masses $\sim 10^9 M_\odot$:
\begin{equation}
    \begin{aligned}
        \left.\frac{{\rm d}n}{{\rm d}m}\right|_{\mathrm{PBH}} = \frac{A}{\sqrt{2\pi}\sigma m} \exp\left(\frac{(\log_{10}m - \mu)^2}{\sigma^2}\right).
    \end{aligned}
\end{equation}
Here, we adopt $\mu=9$ and $\sigma=0.05$. The normalization factor $A$
is fixed by the comoving number density $n_{\rm PBH}$.

\begin{figure}[!tbp]
  \centering
  \includegraphics[width=0.68\columnwidth]{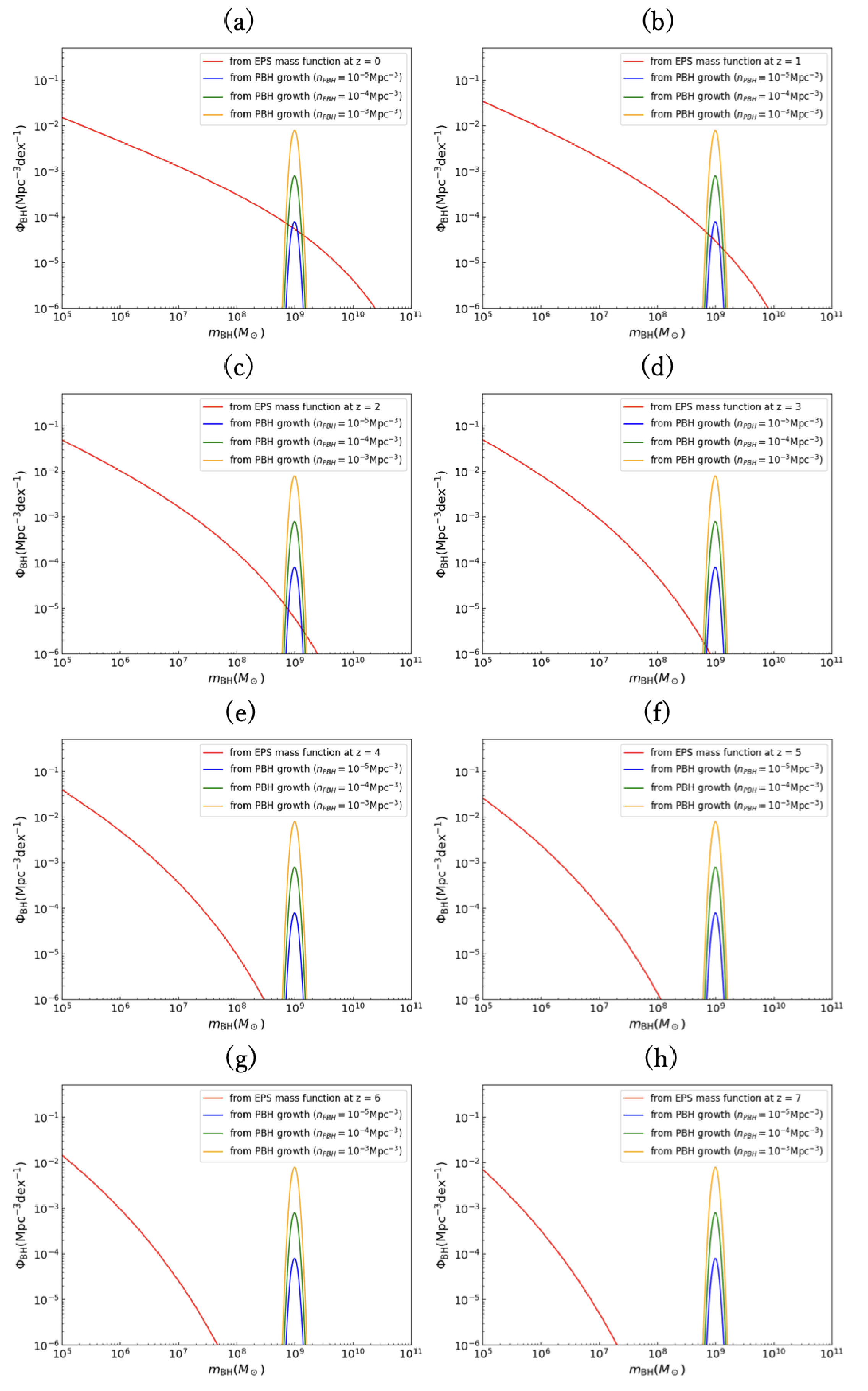}
  \caption{BH mass function given by Eq.~(\ref{eq:fullBHMF}) at
      different redshifts. The horizontal axis shows the SMBH mass
      ($M_\odot$), and the vertical axis shows the BH mass function
      ($\mathrm{Mpc^{-3}dex^{-1}}$). The red solid line denotes the
      EPS BH mass function, while the blue, green, and yellow lines
      represent the PBH seed contribution with comoving number
      densities of
      $n_{\rm PBH}=10^{-5}, 10^{-4}, 10^{-3} \, \mathrm{Mpc^{-3}}$,
      respectively. The panels show the cases for (a) $z=0$, (b) $z=1$, (c)
      $z=2$, (d) $z=3$, (e) $z=4$, (f) $z=5$, (g) $z=6$, and (h)
      $z=7$.}
    \label{fig:fullBHMF}
\end{figure}

Fig.~\ref{fig:fullBHMF} shows the EPS BH mass function (red solid
line) and the additional contribution from PBH seeds (blue, green, and
yellow lines) at $z$ = 0 -- 7. The blue, green, and yellow curves
correspond to the PBH seed comoving number densities of
$n_{\rm PBH} = 10^{-5}, 10^{-4}, 10^{-3} \, \mathrm{Mpc^{-3}}$,
respectively.  With this formulation, the contribution from the PBH
seeds can be incorporated into the BH merger rate given in
Eq.~(\ref{eq:BH_merger_rate2}). In this study, we compute the GW
energy density spectrum including this PBH component, and determine
the value of the comoving PBH number density $n_{\rm PBH}$ that
reproduces the strain amplitude of the NANOGrav 15-year.

\section{Results}
As we have discussed in the previous section, in order to fit the
strain amplitude observed by the NANOGrav 15-year
$A=2.4^{+0.7}_{-0.6} \times10^{-15}$ (median + 90\% credible interval)
at $f = 1 \mathrm{yr}^{-1}$, we find that the required number density
of the PBH seeds is constrained to be
$1.97 \times 10^{-4} \mathrm{Mpc}^{-3} \lesssim n_{\rm PBH} \lesssim
6.16 \times 10^{-4} \mathrm{Mpc}^{-3}$.

\begin{figure}[!h]
  \centering
  \includegraphics[width=1.0\columnwidth]{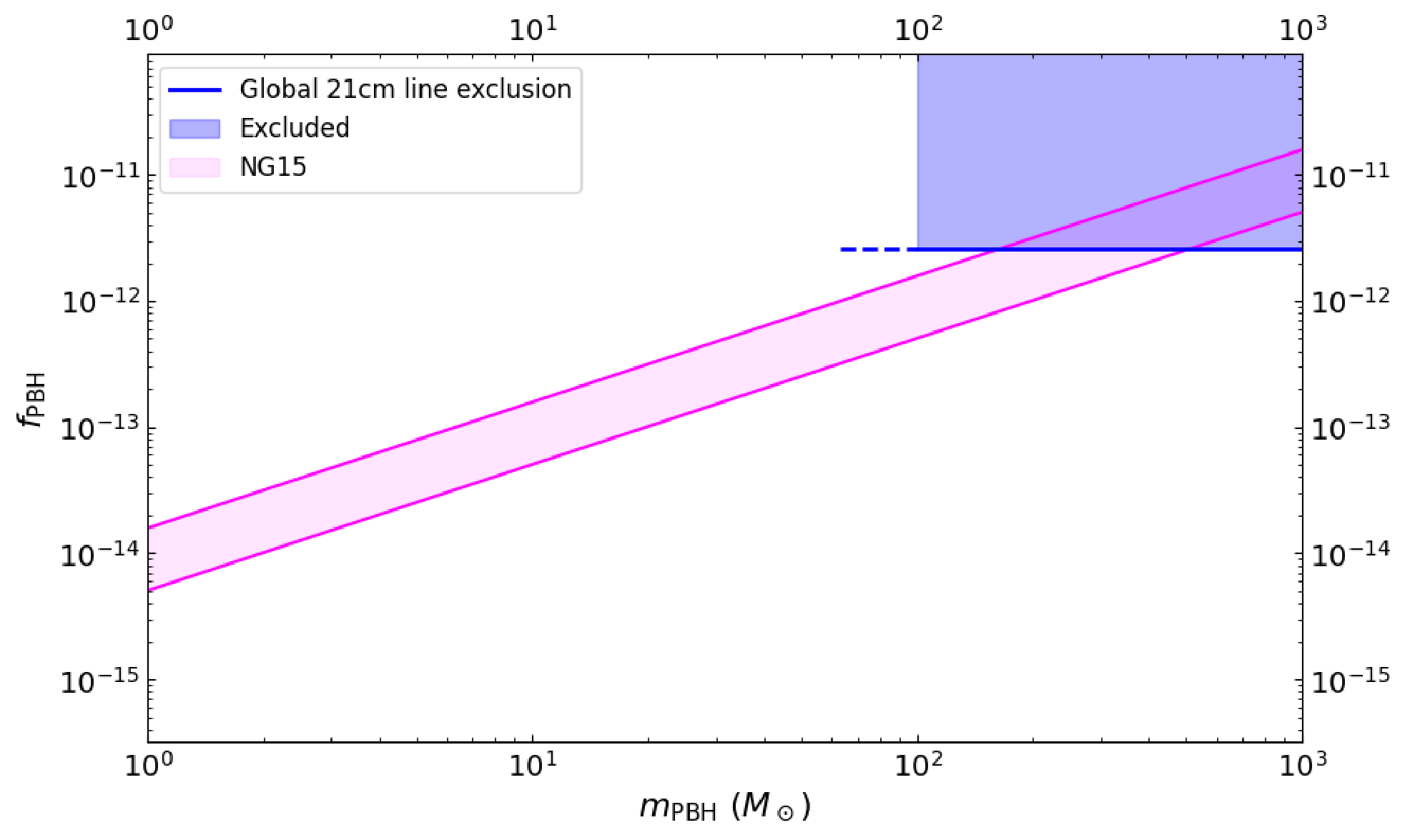}
  \caption{
    Abundance of the PBH seeds required to fit the GW signal
    of the NANOGrav 15-year. The horizontal axis shows the PBH mass
    ($m_{\rm PBH}$) in $M_{\odot}$, while the vertical axis shows
    the fraction of the energy density of the PBHs to the energy density of the 
    CDM ($f_{\rm PBH}$). The oblique magenta band represents the
    allowed region of $f_{\rm PBH}$ for each PBH mass. The blue
    shaded area denotes the region excluded by the observations of the cosmological
    high-redshifted global 21cm line, 
    assuming the growth of the PBHs via accretion to the SMBHs until 
    $z\sim7$.
  }
  \label{fig:f_PBH_result}
\end{figure}

Then, the energy fraction of the seed PBHs to the CDM at present is
given by
\begin{equation}
\begin{aligned}
f_{\rm PBH} = \frac{m_{\rm PBH} n_{\rm PBH} }{\rho_{\rm CDM,0}},
\end{aligned}
\end{equation}
with the energy density of the CDM $\rho_{\rm CDM,0}$ at present.
Using this relation, we derived the allowed region of $f_{\rm PBH}$ to
fit the signal of the NANOGrav 15-year, as shown in
Fig.~\ref{fig:f_PBH_result}.

The blue shaded region indicates the excluded region derived from the
observations of the high-redshifted cosmological global 21cm line
based on the accretion on to the PBHs~\cite{Kohri:2022wzp}. According
to these constraints, if the PBHs with a number density of
$n_{\rm PBH}=10^{-5} \mathrm{Mpc}^{-3}$ grow to $10^9 M_\odot$ until
the redshift $z~\sim 7$, the initial mass must satisfy
$M_{\rm PBH}\lesssim10^4 M_\odot$. Similarly, for
$n_{\rm PBH}=10^{-4} \mathrm{Mpc}^{-3}$, the constraint is
$M_{\rm PBH}\lesssim10^3 M_\odot$, and for
$n_{\rm PBH}=10^{-3} \mathrm{Mpc}^{-3}$, the constraint becomes
$M_{\rm PBH}\lesssim10^2 M_\odot$. These bounds imply that PBH
abundances with $f_{\rm PBH} \geq 2.5 \times 10^{-12}$ are excluded.
It should be noted, however, that this constraint applies when the
PBHs reside in environments where the accretion is efficient which is
similar to the standard scenario for the evolution of the SMBHs from a
seed BH. Furthermore, following~\cite{Kohri:2022wzp}, we restrict the
discussion to the range
$10^{-5} \mathrm{Mpc}^{-3} \leq n_{\rm PBH} \leq 10^{-3}
\mathrm{Mpc}^{-3}$,
and do not examine whether similar limits apply outside this interval.

In addition, within the range of the PBH masses allowed by the 21cm
observations, $1 M_\odot \lesssim m_{\rm PBH} \lesssim 10^3 M_\odot$,
we do not show constraints from the other probes~\cite{Carr:2020gox}
such as the $\mu-$distortion of the CMB~\cite{Kohri:2014lza}, the
accoustic rehating~\cite{Jeong:2014gna,Inomata:2016uip} or the CMB
polarization~\cite{Poulin:2017bwe,Serpico:2020ehh} because they do not
affect the parameter space considered here.

The magenta band in Fig.~\ref{fig:f_PBH_result} represents the allowed
region of $f_{\rm PBH}$ corresponding to the signal of the NANOGrav
15-year. The upper bound of this region corresponds to the upper value
of the observed strain amplitude, while the lower bound corresponds to
its lower value. We find that an abundance of the seed PBHs in the range
$10^{-14} \lesssim f_{\rm PBH} \lesssim 10^{-12}$
can fit the GW signal of the NANOGrav 15-year.  These results
demonstrate that even a small additional population of the PBH seeds
is capable of enhancing the amplitude to the level
observed by NANOGrav 15-year after its growth to the SMBH until $z \sim 7$
by the Eddington or super-Eddington accretion on to the seed PBH.

The accretion disk model adopted in this work assumes, as in Refs.\cite{Poulin:2017bwe,Serpico:2020ehh}, either a standard disk or an advective cooled disk.
In our model, the required number density of seed black holes is very small. Under a natural setup, the patchy heating due to accretion, which could affect the CMB through EE polarization, is not detectable with current observations by the Planck satellite. Therefore, the required number density is significantly below the observational upper bounds derived in Refs.\cite{Poulin:2017bwe,Serpico:2020ehh}, and is thus allowed.

To produce the seed PBHs with the masses of $1 M_{\odot}$ -- $10^3
M_{\odot}$ by the gravitational collpase of large curvature
perturbation at small scales, the scalar-induced gravitation wave
(SIGW) with the frequencies of ${\cal O}(0.1) -- {\cal O}(1)$ nHz can
be simultaneously produced. However, the amplitude of the SIGW
generated in conjunction with the seed PBHs is too small to be
observed by NANOGrav. In future, the IPTA or SKA will observe
signatures of the SIGWs to test this scenario. Veryfinng this
possibility is beyond the scope of this work. Thus, we will report it
in a separate paper.

\section{Conclusion}

In this paper, we have investigated theoretical models that fit the
stochastic GWB in the nHz band reported by the NANOGrav
15-year observation.  In astrophysics, it is known that the nHz GWs
are generated by the merger of binary SMBHs accompanying galaxy
mergers, with each SMBH residing at the center of its galaxy.
However, even the latest numerical simulations indicate that the
theoretical GW signal from such astrophysical binary SMBH mergers is
much smaller than the GW signal reported by the NANOGrav 15-year
observation.

Therefore, we have studied the possibility of a GWB signal
from the GWs generated by the merger of the other type of binary
SMBHs, evolved from the PBHs as the seed BHs predicted in the
new-physics model beyond the standard model. As a result, by following
a completely parallel line of reasoning to the astrophysical scenario,
we successfully fitted the GWs reported by NANOGrav 15-year from the
mergers of the binary SMBHs originated from the seed PBHs. In this
scenario, most of the SMBHs could have been evolved from the seed PBHs
through the high mass-accretion rate. Then, the model parameters are
specifically obtained as follows. The fraction of the PBHs to the CDM
can be $10^{-14} \lesssim f_{\rm PBH} \lesssim 10^{-12}$ with the
masses of the seed PBHs,
$1 M_{\odot} \lesssim m_{\rm PBH} \lesssim 10^{3} M_{\odot}$.

However, such high mass-accretion rate at Eddington or super-Eddington
rates, necessary for successful growth from the seed PBHs to the SMBHs,
should have affected the cosmological high-redshift 21cm line
emission. Nevertheless, the seed PBHs with an abundance
$f_{\rm PBH} < 2.5 \times 10^{-12}$, which is independent of the mass
of the seed PBHs, can grow into SMBHs without conflicting with the
existing observations of the cosmological high-redshift 21cm
line. Conversely, we conclude that more precise 21cm line observations
in future, e.g., the phase 2 of the Square Kilometer Array
(SKA)~\cite{Weltman:2018zrl}, will be able to test this scenario.  By
measuring the cosmological high-redshift 21cm line emission, we will
find the system of the accretion disks around the seed PBHs on the way
to evolve to the high-redshift SMBHs, which can be the source for the
observed nHz GW. In addition, we propose that the transitional phase in which black holes grow into the SMBHs via accretion onto the PBH seeds may be observed as Little Red Dots. By confirming that the seed BHs evolving to the
high-redshift SMBHs should be the PBHs, we can obtain a hint for
distinguishing theoretical models of primordial inflation through such
future observations.

\begin{acknowledgments}
This work was in part supported by JSPS KAKENHI Grants
Nos. JP23KF0289, JP24K07027, and MEXT KAKENHI Grants No. JP24H01825.
\end{acknowledgments}

\bibliography{references}

\end{document}